\documentclass[twocolumn,showpacs,preprintnumbers,amsmath,amssymb]{revtex4}
\usepackage{graphicx,color}
\usepackage{bm}

\begin{document}

\title{An Explanation for Heavy Quark Energy Loss Puzzle by Flow Effects}

\author{Luan Cheng$^{a,b}$ and Enke Wang$^{a,b}$}

\affiliation{ $^a$Institute of Particle Physics, Huazhong Normal
University, Wuhan 430079, China\\
$^b$Key Laboratory of Quark $\&$ Lepton Physics (Huazhong Normal
University), Ministry of Education, Wuhan 430079, China}



\begin{abstract}
The heavy quark energy loss puzzle is explained by collective flow
effects in a dynamic medium. The dead cone and LPM effect are found
to be changed comparing to the static medium case. Instead of only
one dead cone in the static medium, the collective flow induces two
dead cones from two different kinds of processes. One is from the
projectile emitting gluon process, the same as that in the static
medium. The other is from the gluon emission off the exchanged gluon
process, decreasing with increasing flow velocity $v_z$ along jet
direction, which lead to the increase of heavy quark energy loss.
The differences of the effective average energy loss among charm,
bottom and light quarks are very little from a full 3D ideal
hydrodynamic simulation for 0-10$\%$ central Au-Au collisions at
RHIC energy. This would yield similar high $p_{T}$ suppressions
between light and heavy quarks for central Au-Au collisions.


\end{abstract}

\pacs{12.38.Mh, 24.85.+p; 25.75.Bh; 25.75.-q}

\maketitle

{\it Introduction} --- Jet quenching \cite{WG1992} or suppression of
large transverse momentum ($p_{T}$) hadrons caused by energy loss of
propagating partons in dense medium has become a powerful tool in
studying the properties of quark-gluon plasma (QGP) in high energy
heavy ion physics. Light quark and gluon jet quenching observed  at
the relativistic heavy ion collider (RHIC) are remarkably consistent
thus far with predictions \cite{GVWZ,GLV1,VG,Wang2} by jet quenching
theory. However, recent non-photonic single electron data
\cite{Adler,Abelev}, which present an indirect probe of heavy quark
energy loss, have significantly challenged the scenario of jet
quenching theory in a static medium. It is estimated that the heavy
quark mass leads to a kinematic ``dead cone'' effect that reduces
significantly the induced radiative energy loss of heavy
quarks\cite{Djordjevic,Armestro,Dokshitzer,Zhang}, but experimental
data shows that
in the $p_{T}= 4-8$ GeV region a suppression of electrons similar to
that of light hadrons for central Au-Au
collisions\cite{Adler,Abelev}, much larger than predicted.

In the majority of currently available studies the medium-induced
radiative heavy quark energy loss is computed by assuming that the
QCD medium consists of randomly distributed static scattering
centers with no energy and little momentum
transfer\cite{Djordjevic}. Later, radiative energy loss in a finite
dynamical QCD medium is studied and is found that the consideration
of dynamic medium improves the agreement with available data, but
still does not yield a perfect description\cite{Heinz1}. All these
computation suffers from one crucial drawback: They are all based on
one assumption that the momentum transfer $q$ is very little, so
that its effect on the dead cone, which is obtained from the
integration of propagators, is little. That induces the same dead
cone in the QGP medium as that in the vacuum. However, Flow velocity
leads to dynamic scattering centers instead of being static, the
influence of the momentum transfer $q$ on propagator of exchanged
gluon is very large. That will surely change the dead cone and heavy
quark energy loss.

\begin{figure}
\vspace{-45pt}
\includegraphics[width=80mm]{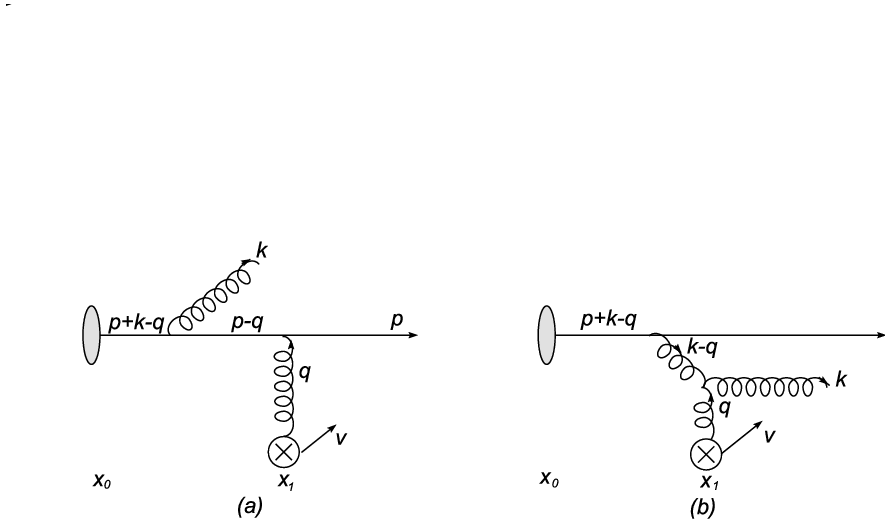}
\vspace{-15pt} \caption{\label{fig:feydia}
  Two kinds of Feynman diagrams contribute to heavy quark energy loss.
  See text for discussion.}
\vspace{-15pt}
\end{figure}

In this letter, we will report a first study of flow effect on dead
cone and LPM effect of heavy quarks, and explain the heavy quark
energy loss puzzle. There are two dead cones of heavy quarks from
two kinds of processes, one is from process I, the gluon emission
from projectile process, such as shown in Fig.\ref{fig:feydia}(a),
the other is from process II, the gluon emission off the exchanged
gluon process resulting from QCD non-Abelian property, such as shown
in Fig.\ref{fig:feydia}(b). The only one dead cone in a static
medium is a particular case that the two dead cones are equal to
each other when flow velocity goes to zero. The dead cone from
process I in the presence of collective flow is the same as that in
the static medium. However, the dead cone from process II is found
to reduce significantly with flow velocity along the jet direction
$v_z$ comparing to the static medium case, leading to an increased
energy loss by heavy quarks. Based on our previous new model
potential \cite{Cheng} with collective flow, using the full 3D ideal
hydrodynamic simulations \cite{Hirano}, we obtain the effective
average energy loss of heavy quark jets, which is nearly the same as
the light quarks'. This agrees with the experimental data at RHIC
where the suppression of non-photonic electrons is similar to that
of light hadrons for central collisions.

{\it Dead Cone in a Dynamic Medium} --- In the observer's system
frame $\Sigma$ the dynamic scattering center moves with flow
velocity $\mathbf{v}$. Consider a hard heavy quark jet with mass $M$
and initial energy $E$ produced at
$\mathbf{x}_0=(z_0,\mathbf{x}_{0\perp})$. It interacts with the
target parton at $\mathbf{x}_1=(z_1,\mathbf{x}_{1\perp})$ with flow
velocity by exchanging gluon with four-momentum $q_i$, radiates a
gluon with four-momentum $k$ and polarization $\epsilon(k)$, and
emerges with final four-momentum $p$. In the rest frame
$\Sigma^{\prime}$ the scattering center is fixed, the momentum
transfer is $q^{\prime}$. With Lorenz boost, one finds
\vspace{-0.08in}
\begin{eqnarray} \left\{
\begin{array}{ll}
\label{q0} q_{i}^{0\prime}= \gamma(q_{i}^0-\mathbf{v} \cdot
\mathbf{q})\, ,
\\
\label{qvector} \mathbf{q}^{\prime}=
\mathbf{q}+\frac{\gamma-1}{\mathbf{v}^2}(\mathbf{v} \cdot
\mathbf{q})\mathbf{v}-\gamma q_i^0 \mathbf{v} \, ,
\end{array}
\right.
\end{eqnarray}
{\vskip -0.08in}
\noindent where $q_i=(q_i^0, \mathbf{q})$ is the
momentum transfer in observer's system frame $\Sigma$. The
collective flow leads to non-zero energy transfer
$q_i^0=\mathbf{v}\cdot\mathbf{q}$ from the target parton to the jet
which differs with $q_i^0=0$ in the static medium from
Gyulassy-Wang's static potential model \cite{WG1992}.

In Ref. \cite{Djordjevic}, it is shown that gluons in the medium can
be approximated as massive transverse plasmons with mass $m_g\approx
\mu/\sqrt{2}$, $\mu$ the Debye screening mass. Then $p$, $k$ and
polarization $\epsilon(k)$ can be written in the light-cone
components, \vspace{-0.08in}
\begin{eqnarray}
 k &=
&[2\omega,\frac{\mathbf{k}_{\perp}^2{+}m_g^2}{2\omega},
  \mathbf{k}_{\perp}]\, ,\quad
 \epsilon(k) = [0,2\frac{\mathbf{\epsilon}_{\perp}{\cdot}
 \mathbf{k}_{\perp}}{xE^+},\mathbf{\epsilon}_{\perp}]\, ,\nonumber \\
p &=& [(1{-}x)E^+{+}2\mathbf{v}{\cdot}
\mathbf{q},\frac{\mathbf{p}_{\perp}^2{+}M^2}{(1{-}x)E^+{+}2\mathbf{v}{\cdot}
\mathbf{q}},\mathbf{p}_{\perp}]\, ,
 \label{lightconecom}
\end{eqnarray}
{\vskip -0.08in}
\noindent where $\omega=xE$, $E^+=2E\gg\mu$.

The ``dead cone" phenomenon, which is induced by mass effect, comes
from heavy quark propagation. Assuming $q$ is little, the dead cone
is obtained by integration of propagators of heavy quarks and
exchanged gluons over $q_z$ for medium-induced radiative energy loss
in the static medium\cite{Djordjevic,Heinz1}. For the gluon emission
from projectile processes such as in Fig.\ref{fig:feydia}(a), the
radiation amplitude can be expressed as $1/({\bf
k}_{\perp}^2+m_g^2+x^2M^2)$. For the the gluon emission off the
exchanged gluon processes such as shown in Fig.\ref{fig:feydia}(b),
the radiation amplitude is $1/(({{\bf k}_{\perp}{-}{\bf
q}_{\perp}})^2+m_g^2+x^2M^2)$. Since ${\bf q}_{\perp}$ is little in
the static medium, the dead cone is the same for both processes
which is equal to
$\theta=\sqrt{m_g^2+x^2M^2}/(xE)$\cite{Djordjevic}.

However, in a dynamic medium, with non-zero energy transfer
$q_i^0=\mathbf{v}\cdot\mathbf{q}$, it is found that the collective
flow changes the poles of the heavy quark propagator with a shifted
term which is dependent on flow velocity.  For process I, the
influence of the momentum transform $q$ on the quark propagator can
be ignored because the transform on q by flow velocity is small
compared to the very hard jet momentum $p$ due to $v<<1$, the dead
cone is the same as that in the static medium, $\theta_1=\theta$.
However, for process II, the influence of momentum transform $q$ on
gluon propagator is significant because the transform on q by flow
velocity is comparable to radiated gluon momentum $k$. The dead cone
is changed to \vspace{-0.08in}
\begin{eqnarray}
  \label{deadcone2}
  \theta_{2}=\frac{ \sqrt{m_g^2+x^2M^2-\frac{2v_z x M^2}{1-v_z}}}{xE}
     \approx  \theta_{1}-v_z\phi ,
\end{eqnarray}
{\vskip -0.08in}
\noindent where the new angle $\phi= M/\omega$. In
this way, the dead cone $\theta_2$ reduces with increasing the flow
velocity along jet direction $v_z$ significantly. This will increase
the contribution of process II to heavy quark energy loss. The only
one dead cone in a static medium is a particular case when $v_z$
goes to zero, then $\theta_2=\theta_1$.

\begin{figure}
\vspace{-10pt}
\includegraphics[width=68mm]{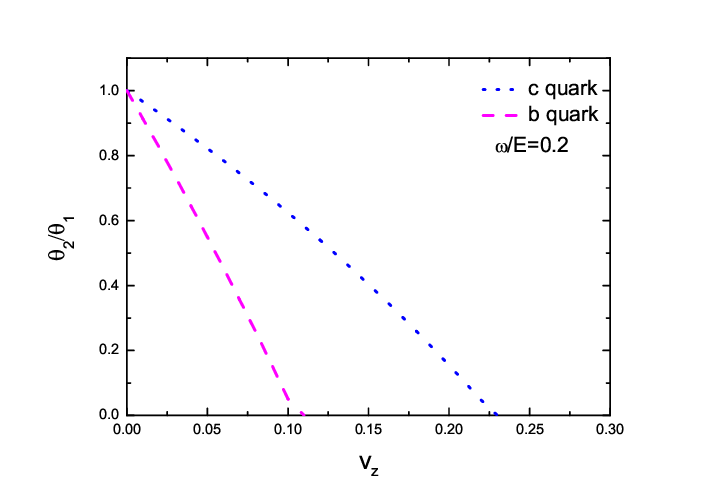}
\vspace{-16pt} \caption{\label{fig:deadcone}
  The ratio $\theta_2/\theta_1$ as function
  of $v_z$
 at $\omega/E=0.2$.} \vspace{-15pt}
\end{figure}

Fig.\ref{fig:deadcone} is the ratio of dead cone $\theta_2$ from
process II to the dead cone $\theta_1$ from process I as function of
$v_z$ at $\omega/E=0.2$. It shows that the dead cone decrease to
zero rapidly with increasing $v_z$. The dead cone of bottom quark
reduces more strongly than that of charm quark because of the
heavier quark mass.

{\it Heavy Quark Energy Loss with Flow} ---
 In our previous work \cite{Cheng} we model the jet
interactions in the presence of collective flow as random color
screened potentials
$A_i^{\nu}(q_i)=(V_i^{flow}(q_i),\mathbf{A}_i^{flow}(q_i))$,
\vspace{-0.08in}
\begin{eqnarray}
 \left\{
\begin{array}{ll}
\label{flowpotential1} V_i^{flow}(q_i)=
2\pi\delta(q_i^0{-}\mathbf{v}{\cdot}\mathbf{q}) e^{-i\mathbf{q}\cdot
\mathbf{x}_i} \tilde{v}(\mathbf{q})T_{a}(j)T_{a}(i)\, ,
\\
\label{flowpotential2} \mathbf{A}_i^{flow}(q_i)=
2\pi\delta(q_i^0{-}\mathbf{v}{\cdot}\mathbf{q}) \mathbf{v}
e^{-i\mathbf{q}\cdot
\mathbf{x}_i}\tilde{v}(\mathbf{q})T_{a}(j)T_{a}(i)\, ,
\end{array}
\right.
\end{eqnarray}
{\vskip -0.08in}
\noindent where
$\tilde{v}(\mathbf{q})={4\pi\alpha_s}/({\mathbf{q}^2-(\mathbf{v}
 \cdot\mathbf{q})^2+\mu^2})$,
 $T_{a}(j)$ and $T_{a}(i)$ the color
matrices for the jet and target parton at position $\mathbf{x}_i$,
respectively. The collective flow changes the color-electric field,
produces a color-magnetic field and leads to a non-zero energy
transfer compared to Gyulassy-Wang's static potential model
\cite{WG1992}.

Here we investigate the rescattering-induced gluon radiation of
heavy quark jet by considering the flow effect resulting from the
moving parton target in a dynamic medium. Opacity is defined as the
mean number of collisions in the medium. We will work in the
framework of opacity expansion as in Ref. \cite{GLV,Wiedua}, which
shows that the first order opacity contribution is dominant because
the higher order corrections contribute little to the radiative
energy loss\cite{GLV}.

To the zeroth order in opacity, the jet has no interaction with the
target parton, that implies that the radiation amplitude and the
dead cone should be the same as that in the static medium in Ref.
\cite{Djordjevic}. To first order in opacity, cross section for the
induced radiation consists of $3^2$ real single scattering and
$2\times4$ double Born contributions in the contact limit. Based on
our new potential with the collective flow in
Eq.(\ref{flowpotential1}), assuming the flow velocity
$|\mathbf{v}|\ll 1$, the medium induced radiation probability to the
first order in opacity can be expressed as \vspace{-0.26in}
\begin{widetext}
\vspace{-0.26in}
\begin{eqnarray}
 \label{probafirstord}
     {{dP^{(1)}}\over d\omega}&=&{{C_2(T)}\over {8\pi d_A d_R}}{N\over A_{\perp}}
     \int {{dx}\over x}  \int
      {{d^2{\bf k}_{\perp}}\over {(2\pi)^2}}\int {{d^2{\bf q}_{\perp}}\over {(2\pi)^2}}
      P({\omega\over E})
[(1{-}v_z)^2  v^2({\bf
     q}_{\perp})]\left\langle
     Tr\left[|R^{(S)}|^2{+}2
      Re\left(R^{(0)\dagger}R^{(D)}\right)\right]\right\rangle
\nonumber\\
     &\approx &\frac{\alpha_s C_R}{2\pi}\frac{L}{l_g} \int {{dx}\over x}  \int
      {d{\bf k}_{\perp}^2} \int {d^2{\bf q}_{\perp}} P(x) |{\bar v}({\bf
      q}_{\perp})|^2(1{-}v_z)^2
       \Bigl((2\mathbf{C}_2^2{-}\mathbf{H}_1{\cdot}
      \mathbf{C}_1{-}\mathbf{H}_1{\cdot}
      \mathbf{H}_2)\left\langle Re(1{-}e^{i\frac{\omega_{1}{+}\omega_{m}{-}\omega_M}
     {1{-}v_z}z_{10}})\right\rangle
\nonumber\\
      &&{+}(\mathbf{H}_1^2{-}\mathbf{H}_1{\cdot}
      \mathbf{C}_1)\left\langle Re(1{-}e^{i\frac{\omega_{1}{+}\omega_{m}}
      {1{-}v_z}z_{10}})\right\rangle
    {+}2v_z\mathbf{H}_1{\cdot} \mathbf{C}_2\left\langle
     Re\Bigl(e^{-i\frac{\omega_M} {1{-}v_z}z_{10}}(1{-}e^{i\frac{\omega_{1}{+}\omega_{m}}
     {1{-}v_z}z_{10}})\Bigr)\right\rangle
\nonumber \\
     &&{+}2v_z(\mathbf{H}_1{\cdot}
     \mathbf{C}_2{-}\mathbf{H}_1^2)\left\langle Re(1{-}e^{i\frac{\omega_{1}{+}\omega_{m}}
     {1{-}v_z}z_{10}})\right\rangle
{-}2v_z\mathbf{H}_1^2\left\langle
     Re(1{-}e^{i\frac{\omega_{1}{+}\omega_{m}} {1{-}v_z}z_{10}})\right\rangle
     \Bigr)\, ,
\end{eqnarray}
\vspace{-0.1in}
\end{widetext}
{\vskip -0.2in} \noindent where $z_{10}=z_1-z_0$, $C_R$ and $C_A$
are the Casimirs of jet in fundamental representation in $d_R$
dimension and the target parton in adjoint representation in $d_A$
dimension, respectively. We will assume all target partons are in
the same $d_T$ dimensional representation with Casimir $C_2(T)$.
$N$, $L$ and $A_{\perp}$ are, respectively, the number, the
thickness, and the transverse area of the targets. $l_g$ is the
mean-free path of gluon. $\alpha_{s}=g^2/4\pi$ the strong coupling
constant and $P_{gq}(x)\equiv P(x)/x=[1+(1-x)^2]/x$ the splitting
function for $q\rightarrow gq$. $|{\bar v}({\bf q}_{\perp})|^2$ is
defined as the normalized distribution of momentum transfer from the
scattering centers as shown in Ref. \cite{Cheng}. The opacity factor
$L/l_g$ reflecting the mean number of rescattering in the medium
arises from the sum over the $N$ distinct targets. In
Eq.(\ref{probafirstord}) we used following quantities,
\vspace{-0.08in}
\begin{eqnarray}
 \label{omega}
   \omega_0&=&\frac{{\bf k}_{\perp}^2}{2\omega}\, ,\quad
   \omega_1=\frac{({\bf k}_{\perp}{-}{\bf q}_{\perp})^2}
      {2\omega}\, ,
 \\
 \label{omegam}
   \omega_m&=&\frac{m_g^2{+}x^2M^2}{2\omega}\, ,\quad
   \omega_M=\frac{v_zx M^2}
      {(1-v_z)\omega}\, ,
 \\
 \label{H1}
   {\bf H}_1&=&\frac{{\bf k}_{\perp}}{{\bf k}_{\perp}^2+m_g^2+x^2M^2}\, ,
 \\
 \label{H2}
   {\bf H}_2&=&\frac{{\bf k}_{\perp}}{{\bf k}_{\perp}^2
   +m_g^2+x^2M^2-\frac{2v_z}{1-v_z}xM^2}\, ,
   \\
 \label{C1}
   {\bf C}_1&=&\frac{{\bf k}_{\perp}{-}{\bf q}_{\perp}}
      {({{\bf k}_{\perp}{-}{\bf q}_{\perp}})^2+m_g^2+x^2M^2}\, ,
 \\
 \label{C2}
   {\bf C}_2&=&\frac{{\bf k}_{\perp}{-}{\bf q}_{\perp}}
      {({{\bf k}_{\perp}{-}{\bf q}_{\perp}})^2+m_g^2+x^2M^2
      -\frac{2v_z}{1-v_z}xM^2}\, .
 \end{eqnarray}
{\vskip -0.08in} \noindent The induced radiation probability in
Eq.(\ref{probafirstord}) depends on flow velocity, different with
the static medium case. The change of dead cone in the the gluon
emission off the exchanged gluon processes can also be read from
Eqs. (\ref{H2}) and (\ref{C2}). Our results agree with the results
 in static medium\cite{Djordjevic} at zero flow velocity.

The gluon formation factor inside $Re(\cdots)$ in
Eq.(\ref{probafirstord}) reflects the destructive interference of
the non-Abelian LPM effect \cite{LPM}. Comparing to the radiated
gluon formation time $\tau_f^{static}=1/(\omega_1+\omega_m)$ in the
static medium in Ref.\cite{Djordjevic}, from the gluon formation
factor in Eq.(\ref{probafirstord}) we see that the radiated gluon
formation time in the medium with collective flow contracts by a
factor $1-v_z$. From Eq. (\ref{probafirstord}) we see clearly three
different formation time,
$\tau_{f1}=(1{-}v_z)/(\omega_1{+}\omega_m)$ is the gluon formation
time contributed by the process I, $
\tau_{f2}=(1{-}v_z)/(\omega_1{+}\omega_m {-}\omega_M)$ is that from
process II, $\tau_{f3}=(1{-}v_z)/{\omega_M}$ comes from the
interference terms of process I and process II.
The gluon formation factor should be averaged over the longitudinal
target profile which is defined as $\left\langle
\cdots\right\rangle=\int dz \rho(z)\cdots$. We take the target
distribution as an exponential Gaussian form
$\rho(z)=\exp(-z/L_e)/L_e$ with $L_e=L/2$.

\begin{figure}
\vspace{-10pt}
\includegraphics[width=68mm]{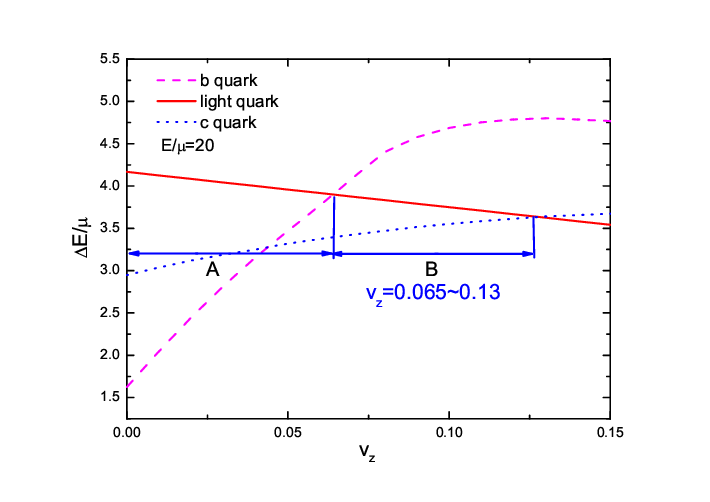}
\vspace{-20pt} \caption{\label{fig:eloss1}
  The energy loss of charm, bottom and light quarks as function of
  $v_z$
    when $E/\mu=20$, $L/l_g=5$.}
\vspace{-20pt}
\end{figure}

From Eq. (\ref{probafirstord}) we obtain the induced energy loss to
the first order of the opacity as $\Delta E =\int d\omega\omega
dP^{(1)}/d\omega$. Its numerical results are shown in Fig.
\ref{fig:eloss1} for c, b and light quarks as $E/\mu=20$, $L/l_g=5$.
At zero flow velocity, the light quark energy loss is larger than c
and b quark energy loss because of the dead cone effect for heavy
quark jets as shown in Ref. \cite{Djordjevic} in the static medium.
However, the heavy quark energy loss increases with increasing flow
velocity because of the decrease of the dead cone effect due to
collective flow. The heavier the quark mass is, the more rapidly the
heavy quark energy loss increases. The light quark energy loss
decreases with increasing flow velocity as shown in our previous
work \cite{Cheng}. The heavy quark energy loss can be larger than
light quark's because not only the dead cone $\theta_2$ but also the
LPM effect is reduced. As shown in Fig. \ref{fig:eloss1}, in the
region of $B$, $0.065{<}v_z{<}0.13$, the energy loss of light quark
is a bit less than that of b quark but a bit larger than that of c
quark. In this region the average heavy and light quark energy loss
is possible to be nearly the same for central Au-Au collisions at
RHIC. However, in region A, light quark energy loss is still larger
than heavy quarks' because of less flow velocity.

For A-A collisions at impact parameter ${\bf b}$, with respect to
collision number the average $v_z$ of rescattering centers in the
QGP medium before freezeout can be expressed as\vspace{-0.08in}
 \begin{equation}
 \label{vz}
  \langle v_z \rangle {=}\frac{\int d^2 \mathbf{r} \int \frac{d \varphi}
  {2\pi}\int d\tau
  \bigl(v_{x}\cos\varphi{+}v_{y}\sin\varphi \bigr)
  \rho\sigma t_At_B}
  {\int d^2 \mathbf{r} \int d\tau \rho\sigma t_At_B}\, ,
 \end{equation}
{\vskip -0.08in}
\noindent where $\varphi$ is the angle between jet
and $x$ axis,
$v_{x}=v_{x}(|\mathbf{r}+\mathbf{n}\tau-\mathbf{b}/2|)$,
$v_{y}=v_{y}(|\mathbf{r}+\mathbf{n}\tau-\mathbf{b}/2|)$ is the
transverse flow velocity of the expanding elliptic medium along the
minor and major semi-axes, $\mathbf{n}$ is the unit vector along jet
direction. Parton density
$\rho=\rho(\tau,\mathbf{b},\mathbf{r}+\mathbf{n}\tau)$, nuclear
thickness functions $t_A=t_A(|\mathbf{r}|)$,
$t_B=t_B(|\mathbf{r}-\mathbf{b}|)$. Cross section $\sigma=C_a
2\pi\alpha_s^2/\mu^2$ ($C_a=1$ for $qg$ and 9/4 for $gg$ scattering)
obtained in pQCD\cite{Wang2}, $\mu=g_sT(\tau)$ at temperature
$T(\tau)$. Using the data from a full 3D ideal hydrodynamic
simulations\cite{Hirano}, we obtain $\langle v_z\rangle=0.08$ for
$0-10\%$ central events of Au-Au collisions at RHIC energy, which
lies in region $B$ as shown in Fig.\ref{fig:eloss1}; $\langle
v_z\rangle=0.016$ for $40-60\%$ central events of Au-Au collisions
at RHIC energy, which lies in region $A$ as shown in
Fig.\ref{fig:eloss1}.

\begin{figure}
\vspace{-20pt}
\begin{center}
\includegraphics[width=90mm,height=60mm]{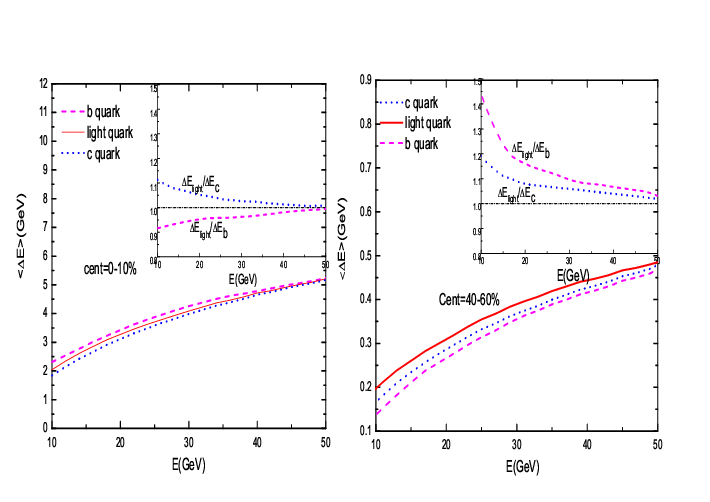}
\end{center}
\vspace{-30pt} \caption{\label{fig:eloss9} The effective average
energy loss of c, b and light quarks as function of jet energy $E$
in the presence of collective flow for $0-10\%$ and $40-60\%$
central events of Au-Au collisions at RHIC energy. Inserted box:
energy loss ratio $\Delta E_{light}/\Delta E_{c}$ and $\Delta
E_{light}/\Delta E_{b}$.} \vspace{-20pt}
\end{figure}

The effective average energy loss of parton jet for A-A collisions
can be written as
\vspace{-0.08in}
 \begin{equation}
 \label{elossaverage}
  \langle \Delta E \rangle =\frac{\int d^2 \mathbf{r} \int \frac{d \phi}{2\pi}\int d\tau \int d\omega
  \omega {{dP^{(1)}}\over d\omega}
  t_At_B}{\int d^2 \mathbf{r} \int d\tau t_At_B}\, .
 \end{equation}
{\vskip -0.08in}
\noindent
Fig.\ref{fig:eloss9} is the effective average energy loss of c, b
and light quarks as function of jet energy $E$ in the presence of
collective flow for $0-10\%$ and $40-60\%$ central events of Au-Au
collisions at RHIC energy. Shown in the inserted box is $\Delta
E_{light}/\Delta E_{c}$, the ratio of effective average energy loss
between light quark's and c quark's, and $\Delta E_{light}/\Delta
E_{b}$, that between light quark's and b quark's. It is shown that
for $0-10\%$ central events, the effective average energy loss of
light quark is a bit less than that of b quark, but a bit larger
than that of c quark. The difference of the effective average energy
loss among three quarks is very little, which implies that the light
and heavy quarks have almost the same suppression of high $p_T$
hadron spectrum for Au-Au collisions at RHIC energy. For $40-60\%$
central events of Au-Au collisions at RHIC energy, the average flow
velocity is less, so that the light quark energy loss is larger than
c and b quark's.

{\it Conclusion} --- In summary, considering the flow effect in a
dynamic medium, we studied ``dead cone" and LPM effect, and give an
explanation for the heavy quark energy loss puzzle. There are two
dead cones for two kinds of processes. The only one dead cone in a
static medium is a particular case with zero flow velocity. The dead
cone $\theta_2$ decreases with increasing $v_z$, leading to the
increase of the contribution from the gluon emission off the
exchanged gluon processes and heavy quark energy loss.
 It has been
shown that in the region of $0.065<v_z<0.13$, the light and heavy
quark energy loss are nearly the same.  For $0-10\%$ central events
of Au-Au collisions at RHIC energy, by using the data from 3D ideal
hydrodynamic simulations we obtain that the average velocity of
rescattering centers along the jet direction $\langle
v_z\rangle$=0.08, the difference of the effective average energy
loss among the charm, bottom and light quarks is very little. But
for $40-60\%$ central events, $\langle v_z\rangle$=0.016, light
quark energy loss is larger than heavy quark's. Our results shall
have implications for comparisons between theory and experiment in
the future.

We thank Fuqiang Wang, Xin-Nian Wang, M. Gyulassy and Xiao-Fang Chen
for helpful discussions. This work was supported by NSFC of China
under Projects No. 10825523, No. 10635020 and No. 10875052, by MOE
of China under Projects No. IRT0624, by MOST of China under Project
No. 2008CB317106; and by MOE and SAFEA of China under Project No.
PITDU-B08033.

\end{document}